\documentclass[submission,copyright,creativecommons,noncommercial]{eptcs}
\providecommand{\event}{LSFA 2022} 
\usepackage{underscore}
\usepackage{amsmath}
\usepackage{amsthm}
\usepackage{amssymb}
\usepackage{stmaryrd}
\usepackage[frozencache,cachedir=minted-cache]{minted}
\setminted{linenos=false,fontfamily=tt}
\newtheorem{definition}{Definition}
\newtheorem{lemma}{Lemma}
\newtheorem{theorem}{Theorem}
\newtheorem{corollary}{Corollary}
\newtheorem{example}{Example}
\usepackage{zi4}
\newcommand{\supp}[1]{\mathrm{supp}(#1)}
\newcommand{\lastEl}[1]{\mathit{last}(#1)}
\newcommand{\tofpUn}[1]{\mathit{toFP}(#1)}
\newcommand{\tofpUnS}[1]{\mathit{toFP}^{*}(#1)}
\newcommand{\tofp}[2]{\mathit{toFP}(#1,#2)}
\newcommand{\fromFP}[1]{\llbracket #1 \rrbracket}
\newcommand{\swap}[2]{(#1\;#2)}
\newcommand{\cycle}[3]{\mathit{cycle}_{#1}(#2,#3)}
\newcommand{\cycles}[4]{\mathit{cycles}_{#1}(#2,#3,#4)}

\newcommand{\tuple}[2]{\langle #1 , #2 \rangle}

\newcommand{\gact}[2]{#1 \bullet #2}

\newcommand{\gactUn}{\gact{\_}{\_}}

\newcommand{\Sym}[1]{\mathit{Sym}(#1)}
\newcommand{\SymA}{\Sym{\mathbb{A}}}
\newcommand{\Perm}[1]{\mathit{Perm}(#1)}
\newcommand{\PermA}{\Perm{\mathbb{A}}}

\newcommand{\agdainline}[1]{\mintinline{agda}{#1}}

\DeclareUnicodeCharacter{03B4}{\ensuremath{\delta}}
\DeclareUnicodeCharacter{03B8}{\ensuremath{\theta}}
\DeclareUnicodeCharacter{03BB}{\ensuremath{\lambda}}
\DeclareUnicodeCharacter{03BC}{\ensuremath{\mu}}
\DeclareUnicodeCharacter{03BC}{\ensuremath{\pi}}
\DeclareUnicodeCharacter{03C3}{\ensuremath{\sigma}}
\DeclareUnicodeCharacter{03C0}{\ensuremath{\pi}}
\DeclareUnicodeCharacter{03D5}{\ensuremath{\phi}}
\DeclareUnicodeCharacter{2200}{\ensuremath{\forall}}
\DeclareUnicodeCharacter{2237}{\ensuremath{::}}
\DeclareUnicodeCharacter{2115}{\ensuremath{\mathbb{N}}}
\DeclareUnicodeCharacter{2248}{\ensuremath{\approx}}
\DeclareUnicodeCharacter{2219}{\ensuremath{\cdot}}
\DeclareUnicodeCharacter{2090}{\ensuremath{_{\text{a}}}}
\DeclareUnicodeCharacter{220E}{\ensuremath{\blacksquare}}
\DeclareUnicodeCharacter{2286}{\ensuremath{\subseteq}}
\DeclareUnicodeCharacter{2287}{\ensuremath{\supseteq}}
\DeclareUnicodeCharacter{2208}{\ensuremath{\in}}
\DeclareUnicodeCharacter{2209}{\ensuremath{\notin}}

\DeclareUnicodeCharacter{2080}{\ensuremath{_0}}
\DeclareUnicodeCharacter{2081}{\ensuremath{_{1}}}
\DeclareUnicodeCharacter{2082}{\ensuremath{_{2}}}
\DeclareUnicodeCharacter{2083}{\ensuremath{_{3}}}
\DeclareUnicodeCharacter{2084}{\ensuremath{_4}}

\DeclareUnicodeCharacter{2113}{\ensuremath{\ell}}
\DeclareUnicodeCharacter{2294}{\ensuremath{\sqcup}}
\DeclareUnicodeCharacter{2261}{\ensuremath{\equiv}}
\DeclareUnicodeCharacter{207A}{\ensuremath{^{\text{+}}}}
\DeclareUnicodeCharacter{02B3}{\ensuremath{^{\text{r}}}}
\DeclareUnicodeCharacter{02E1}{\ensuremath{^{\text{1}}}}
\DeclareUnicodeCharacter{2249}{\ensuremath{\not\approx}}
\DeclareUnicodeCharacter{209B}{\ensuremath{_s}}
\DeclareUnicodeCharacter{03B5}{\ensuremath{\epsilon}}
\DeclareUnicodeCharacter{207B}{\ensuremath{^{-}}}
\DeclareUnicodeCharacter{2032}{\ensuremath{'}}
\DeclareUnicodeCharacter{209A}{\ensuremath{_{p}}}
\DeclareUnicodeCharacter{2218}{\ensuremath{\circ}}
\DeclareUnicodeCharacter{27E6}{\ensuremath{\llbracket}}
\DeclareUnicodeCharacter{27E7}{\ensuremath{\rrbracket}}
\DeclareUnicodeCharacter{225F}{\ensuremath{\overset{?}{=}}}
\DeclareUnicodeCharacter{03A3}{\ensuremath{\Sigma}}
\DeclareUnicodeCharacter{03B9}{\ensuremath{\iota}}
\DeclareUnicodeCharacter{0394}{\ensuremath{\Delta}}
\DeclareUnicodeCharacter{022A5}{\ensuremath{\bot}}
\DeclareUnicodeCharacter{21D2}{\ensuremath{\Rightarrow}}
\DeclareUnicodeCharacter{25CF}{\ensuremath{\bullet}}
\DeclareUnicodeCharacter{03C1}{\ensuremath{\rho}}
\DeclareUnicodeCharacter{2264}{\ensuremath{\leqslant}}

\DeclareUnicodeCharacter{FFFD}{\ensuremath{\mathbb{A}}}
\usepackage{tikz}
\usetikzlibrary{positioning}

\title{Nominal Sets in Agda\\
  A Fresh and Immature Mechanization}
\author{
  Miguel Pagano\thanks{Most of this work was done in a research leave in ORT Uruguay, financed by Agencia Nacional de Investigación e Innovación (ANII) of Uruguay.}
  \institute{FAMAF - Universidad Nacional de Córdoba\\ Córdoba, Argentina}
  \email{miguel.pagano@unc.edu.ar}
  \and
  José E. Solsona
  \institute{Facultad de Ingeniería - Universidad ORT Uruguay\\
    Montevideo, Uruguay}
  \email{solsona@ort.edu.uy}
}
\def\titlerunning{Nominal Sets in Agda}
\def\authorrunning{M. Pagano and J. E. Solsona}
\begin{document}
\maketitle

\begin{abstract}
  In this paper we present our current development on a new
  formalization of nominal sets in Agda. Our first motivation in
  having another formalization was to understand better nominal sets
  and to have a playground for testing type systems based on nominal
  logic. Not surprisingly, we have independently built up the same
  hierarchy of types leading to nominal sets. We diverge from other
  formalizations in how to conceive finite permutations: in our
  formalization a finite permutation is a permutation (i.e. a
  bijection) whose domain is finite. Finite permutations have
  different representations, for instance as compositions of
  transpositions (the predominant in other formalizations) or
  compositions of disjoint cycles.  We prove that these
  representations are equivalent and use them to normalize (up to
  composition order of independent transpositions) compositions of
  transpositions.
\end{abstract}

\section{Introduction}
\label{sec:intro}

Nominal sets were introduced to Computer Science by Gabbay and Pitts
to give an adequate mathematical universe that permits the definition
of inductive sets with binding \cite{Gabbay2002}. Instead of taking
equivalence classes of inductively defined sets (as in a formal
treatment of, say, the Lambda Calculus) or a particular representation
of the variables (as in the de Bruijn approach to Lambda Calculus),
nominal sets have a notion of name abstraction that ensures all the
properties expected for binders; in particular, alpha-equivalent
lambda terms are represented by the same element of the nominal set of
lambda terms.

In this paper we present a new mechanization \cite{pagano2022} of
nominal sets.  Most of the current mechanizations of nominal sets
represent finite permutations as compositions of transpositions, where
transpositions are represented by pairs of atoms and compositions as
lists. In contrast, our starting point is permutations (i.e. bijective
functions); finite permutations are permutations that can be
represented by composition of transpositions. Moreover they conflate
the set of atoms mentioned in a list with the domain of the
(represented) permutation. Pondering about this issue, we decided to
develop a ``normalization'' procedure for representations of finite
permutations; in order to prove its correctness, we were driven to
introduce a cycle notation.

The rest of this paper is structured into four sections. In
Sect.~\ref{sec:fundamentals} we summarize the fundamentals of Nominal
Sets; in Sect.~\ref{sec:perm} we explain the different representations
of finite permutations and their equivalence; then, in
Sect.~\ref{sec:formalization} we present the most salient aspects of
our mechanization in Agda; and finally in Sect.~\ref{sec:conclusion}
we conclude by mentioning related works and contrasting them with our
approach, indicating also our next steps. We assume some knowledge of
Agda, but also hope that the paper can be followed by someone familiar
with any other language based on type theory.

\section{Fundamentals of Nominal Sets}
\label{sec:fundamentals}
In this section we summarize the main concepts underlying the notion
of Nominal Sets; for a more complete treatment we refer the reader to
\cite{Pitts2013-book}. We repeat the basic definitions of group and
group action. A \emph{group} is a set $G$ with a distinguished element
($\epsilon \in G$, the \emph{unit}), a binary operation
($\_\cdot\_\colon G\times G \to G$, the \emph{multiplication}), and a
unary operation ($\_^{-1} \colon G\to G$, the \emph{inverse}),
satisfying the following axioms:
\begin{align*} 
  \text{Associativity:} && g_1 \cdot (g_2 \cdot g_3) \ &= \ (g_1 \cdot g_2) \cdot g_3 && ,\forall g_1,g_2,g_3 \in G\\ 
  \text{Inverse element:} && g\cdot (g^{-1}) \ &= \ \epsilon \ = \ g^{-1} \cdot g  && ,\forall g \in G\\ 
  \text{Identity element:} && \epsilon\cdot g \ &= \ g \ = \ g\cdot \epsilon && ,\forall g \in G
\end{align*}

\noindent Although a group is given by the tuple
$(G,\epsilon,\_\cdot\_,\_^{-1})$ (and the proofs that these operations
satisfy the axioms) we will refer to the group simply by $G$. A
sub-group of $G$ is a subset $H\subseteq G$ such that $\epsilon\in H$
and $H$ is closed under the inverse and multiplication.

Let $G$ be a group. A \emph{$G$-set} is a set $X$ with an operation
$\gactUn \colon G\times X \to X$ (called the \emph{action})
satisfying:
\begin{align*}
\text{Identity:} && \gact{\epsilon}{x} & \ = \ x                            && ,\forall x \in X\\ 
\text{Compatibility:} && \gact{g_1}{(\gact{g_2}{x})} & \ = \ \gact{(g_1\cdot g_2)}{x}  && ,\forall g_1,g_2 \in G, \forall x \in X
\end{align*}

A morphism between $G$-sets $X$ and $Y$ is a function
$F \colon X \to Y$ that commutes with the actions:
\[
  F\,(\gact{g}{x}) \ = \ \gact{g}{F\,x} %
    \quad\qquad ,\forall g \in G, \forall x \in X
\]
These are called \emph{equivariant} functions. Since $id_{X}$ is
equivariant and the composition of equivariant functions yields an
equivariant function we can talk of the category of $G$-Sets.

Any set $X$ can be seen as a $G$-set by letting $\gact{g}{x} = x$;
such a $G$-set is called the \emph{discrete} $G$-set. Moreover any
group acts on itself by the multiplication.

One can form the (in)finitary product of $G$-sets by defining the
action of $G$ on a tuple in a pointwise manner:
\[
  \gact{g}{\tuple{x_1}{\,x_2}} \ = \
  \tuple{\gact{g}{x_1}}{\,\gact{g}{x_2}}
    \quad\qquad ,\forall g \in G, \forall x_1 \in X_1, \forall x_2 \in X_2
\]
The projections and the product morphism $\tuple{F}{H}$ are
equivariant, assuming that $F$ and $H$ are also
equivariant. $G$-set, as a category, also has co-products.

If $X$ and $Y$ are $G$-sets one can endow the set $Y^X$ of functions
from $X$ to $Y$ with the \emph{conjugate} action:
\[
  (\gact{g}{F})\,x \ = \ \gact{g}{(F\,(\gact{g^{-1}}{x}))}
    \quad\qquad ,\forall g \in G, \forall x \in X \enspace .
\]

\paragraph{$G$-sets over the Permutation Group}
The group of symmetries over a set $X$ consists of $G = \Sym{X}$,
where $\Sym{X}$ is the set of bijections on $X$; the multiplication of
$\Sym{X}$ is composition, the inverse is the inverse bijection, and
the unit is the identity.

Let $\Perm{X}$ be the subset of $\Sym{X}$ of bijections that changes
only finitely many elements; i.e., $f\in\Perm{X}$ if
$\supp{f} = \{x\in X \mid f\,x\not=x \}$ is finite. It is
straightforward to prove that $\Perm{X}$ is a sub-group of
$\Sym{X}$. Of course, if $X$ is finite, then $\Perm{X} = \Sym{X}$.
Notice that $X$ itself is a $\Perm{X}$-set with the action being
function application: $\gact{\pi}{x} \ = \ \mathop{\pi}\,x \enspace$.

In particular, the \emph{transposition} (or \textit{swapping}) of a
pair of elements $x, y \in X$ is the finite permutation
$\swap x y \in \Perm{X}$ given by
\[
\swap x y \, z \ = \ \begin{cases}
y & \text{if } z = x \\
x & \text{if } z = y \\
z & \text{otherwise}
\end{cases} \]

A basic result (in \cite{Hungerford} is proved as Theorem 6.3 and
Corollary 6.5) is that every $\pi \in \Perm{X}$ can be expressed as a
composition of \emph{disjoint cycles}
\[
\pi \ = \ (x_1\ x_2\ \ldots\  x_n) \circ \cdots \circ 
(z_1\ z_2\ \ldots\ z_k)
\]
and every cycle can be expressed as a composition of transpositions
\[
(x_1\ x_2\ \ldots\ x_n)\ = \ \swap{x_1}{x_2} \circ \swap{x_2}{x_3} \circ \cdots \circ \swap{x_{n-1}}{x_n}
\]
Therefore every $\pi \in \Perm{X}$ can be expressed as a composition
of transpositions.  We elaborate on the equivalence of the
representations in Sect.~\ref{sec:perm}.  Let us exhibit this with a
concrete example.

\begin{example}
\label{ex:perm}
Let $f\colon \mathbb{N}\to \mathbb{N}$ be defined as
\begin{align*}
    f\,x \ = \ & \begin{cases}
    (x+2) \,\mathop{mod}\, 6 & \text { if } x \leq 5\\
    x & \text{ else }
    \end{cases}
\end{align*}
Function $f$ is a finite permutation, because it has finite support:
$\{ x \in \mathbb{N} \mid 0 \leq x \leq 5 \}$. Therefore it can be
expressed as the composition of two cycles:
$(1\ 3\ 5)\circ (0\ 2\ 4)$, or alternatively, it can also be expressed
as a composition of four transpositions:
$\swap{1}{3}\circ\swap{3}{5}\circ\swap{0}{2}\circ\swap{2}{4}$.
\end{example}

\paragraph{Nominal Sets}

If we let $X$ be the set of variables for the lambda calculus, then a
permutation on $X$ is a renaming; such a permutation can be lifted to
an action over the set of lambda terms (taking care of the bound
variables). In the nominal parlance one says that $X$ is the set of
\emph{atoms} or that variables are atomic names: an atomic name has no
structure in itself. We only assume that a set of atoms is a countable
infinite set with decidable equality; from now on we will use
$\mathbb{A}$ to refer to a set of atoms.

Let $X$ be a $\PermA$-set. We say that $x \in X$ \emph{is supported by} $A\subseteq\mathbb{A}$
if \[
\forall\, \pi.\ (\forall\, a\in A.\ \pi\,a\,=\,a) \ \implies \ \gact{\pi}{x} \,=\, x \enspace .
\]

\noindent We say that $X$ is a \emph{nominal set} if each element of
$X$ is supported by some finite subset of $\mathbb{A}$. Since each
finite permutation can be decomposed as a composition of
transpositions, then one can prove that the above definition is
equivalent to
\[ \forall\, a,a'\in\mathbb{A}\setminus A.\ \gact{\swap{a}{a'}}{x}
  \,=\, x \enspace .
\]

\noindent The following are some examples of nominal sets:
\begin{itemize}
    \item The discrete $Perm(\mathbb{A})$-set $X$ is nominal, because any $x \in X$ is supported by $\emptyset$. 
    \item $\mathbb{A}$ itself is nominal once equipped with the action
      $\gact{\pi}{a} = \pi\,a$, because any $a\in \mathbb{A}$ is
      supported by $\{a\}$. More in general, any
      $S \subseteq \mathbb{A}$ containing name $a$ is a support for
      $a$.
    \item The set $\lambda$\textit{Term} of $\lambda$-calculus terms, inductively defined by $\ t \ ::= \ V(a) \ \mid \ A(t,t) \ \mid \ L(a,t)\ $ where $a \in \mathbb{A}$, equipped with the action $\gactUn \ \colon Perm(\mathbb{A})\times\lambda\textit{Term} \to \lambda\textit{Term}$ such that
      \begin{align*}
        & \gact{\pi}{V(a)}        \ = \ V(\pi\, a)                    \\ 
        & \gact{\pi}{A(t_1,t_2)}  \ = \ A(\gact{\pi}{t_1},\,\gact{\pi}{t_2})     \\ 
        & \gact{\pi}{L(a,t)}      \ = \ L(\pi\, a,\,\gact{\pi}{t})    
      \end{align*}
      is nominal because any $t \in \lambda$\textit{Term} is supported by $supp(t)=FreeVars(t)$.
\end{itemize}

In his book \cite{Pitts2013-book} Pitts uses classical logic to prove
that if $x$ is supported by some finite set $A$, then there exists a
least supporting set, called \emph{the} support of $x$. As shown by
Swan \cite{Swan2017} one cannot define the least support in a
constructive setting; therefore a formalization in a constructive type
theory should ask for ``some'' finite support. This affects the notion
of freshness: in classical logic we have
\[
x \,\textit{ is fresh for }\, y \ \ \Leftrightarrow \ \  \supp{x} \cap \supp{y} = \emptyset, 
\]
with $x\in X$ and $y\in Y$ being elements of different nominal sets;
but in a constructive setting one has to limit this relation to atoms,
that is
\[
a\in\mathbb{A} \,\textit{ is fresh for }\, x \in X \ \ \Leftrightarrow \ \ a\not\in\supp{x}, 
\]
where $\supp{x}$ is the set supporting $x$, not necessarily the least
one. Notice that the definition is the same (``there exists some
finite support for each element''), but in classical logic that is
sufficient to obtain the least support.

\section{Finite Permutations}
\label{sec:perm}
As we have already said, a finite permutation on a set $A$ can be
explicitly given by:
\begin{enumerate}
    \item a bijection $f : A \to A$ together with its support $\supp{f} \subseteq_{\mathit{fin}} A$; i.e., $a\in \supp{f}$ if and only if $f\,a\not= a$;
    \item a composition of disjoint cycles; concretely, we can think of this as a finite set $R \subseteq_{\mathit{fin}} A^*$ of disjoint cycles, each of them without repeated elements;
    \item a composition of transpositions; that is, a finite sequence of pairs $p : (A\times A)^*$.
\end{enumerate}

We present our proof that these definitions are equivalent. It
basically boils down to define a predicate on sequences of elements in
$A$ not containing repeated elements ensuring that they are cycles for
$f$. We use the usual notation $\swap a b$ to denote the bijection
$\{(a,b),(b,a)\}$.

\begin{definition}[List of transpositions from a cycle] We define $\mathit{toFP}\colon A\times A^* \to (A\times A)^*$.
\[
    \tofp{a}{\rho} = \begin{cases} 
    [] & \text{ if } \rho=[]\\
    (a,b):\tofp{b}{\rho'} & \text{ if } \rho=b:\rho'\\
    \end{cases}
\]
If we know that $\rho=a:\rho'$, then we also write $\tofpUn{\rho}$
to mean $\tofp{a}{\rho'}$.
\end{definition}

\begin{definition}[Permutation from a list of transpositions] Let $as : (A\times A)^*$, then $\fromFP{as} : A \to A$ is defined by recursion on $as$:
\[
    \fromFP{as} = \begin{cases} 
    \mathit{id} & \text{ if } as=[]\\
    \swap{a}{b} \cdot \fromFP{as'} & \text{ if } as =(a,b):as'\\
    \end{cases}
\]
\end{definition}

\begin{definition}[Prefixes]
We say that a non-empty sequence $\rho=[a_1,\ldots,a_n] : A^*$ is a \emph{prefix with head $a_0$} \emph{for} bijection $f$ if:
\begin{enumerate}
    \item $a_0 \in \supp{f}$,
    \item $f\,a_i=a_{i+1}$, and
    \item $a_0\not\in\rho$.
\end{enumerate}
A prefix $\rho$ is \emph{closed} if $f\,a_n = a_0$. Since $\rho$ is non-empty, we denote with $\lastEl{\rho}$ its last element.
\end{definition}

From this simple definition we can deduce:
\begin{lemma}[Properties of prefixes] Let $\rho$ be a prefix with head $a$.
\begin{enumerate}
\item If $\rho'$ is a prefix with head $\lastEl{\rho}$, then its concatenation $\rho\rho'$ is a prefix with head $a$.
    \item $\rho$ has no duplicates.
    \item If $\rho$ is closed and $b\in (a:\rho)$, then $f\,b = \fromFP{\tofp{a}{\rho}}\,b$.
    \item If $b\not\in(a:\rho)$, then $\fromFP{\tofp{a}{\rho}}\,b = b$.
    \end{enumerate}
\end{lemma}

We can extend this definition to a sequence of sequences: let $R = [(a_1,\rho_1),\ldots,(a_m,\rho_m)] : (A\times A^*)^*$, then $R$ is a
list of prefixes, with its head, if each $\rho_i$ is a prefix and $\rho_i\cap \rho_j = \emptyset$.

\begin{lemma}[Correctness of prefixes]
Let $R = [(a_1,\rho_1),\ldots,(a_m,\rho_m)]$ be a list of closed prefixes, then $\fromFP{ \tofp{a_1}{\rho_1}\,\ldots\,\tofp{a_m}{\rho_m}}\,a = f\,a$.
\end{lemma}

This proves that from a representation with cycles one can get
a representation with transpositions. If we can produce a list
of closed prefixes from a finite permutation (as a bijection 
with its support explicitly given) then we have the equivalence.
First we define a function $\mathit{cycle}_f\colon 
\mathbb{N}\times A\to A^*$ such that $\cycle{f}{n}{a}$ computes 
a prefix with head $a$ of length at most $n+1$ by recursion on $n$:
\[
\begin{aligned}
   \cycle{f}{0}{a} &= [f\,a]\\
   \cycle{f}{n+1}{a} &= \begin{cases}
    \rho & \text{ if } f\,b=a\\
    \rho[f\,b] & \text{ otherwise}
   \end{cases}\\
    \text{ where }& \rho = \cycle{f}{n}{a} \text{ and } b = \lastEl{\rho}
\end{aligned}
\]

\noindent We can extend this definition to compute a list of prefixes from a list of atoms:
\[
\begin{aligned}
   \cycles{f}{n}{[]}{R} &= R\\
   \cycles{f}{n}{a:as}{R} &= \begin{cases}
    \cycles{f}{n}{as}{R} & \text{ if } a\in \bigcup R\\
    \cycles{f}{n}{as}{\rho:R} & \text{otherwise}
   \end{cases}\\
   \text{ where }& \rho = a:\cycle{f}{n}{a}
\end{aligned}
\]

\begin{lemma}[Correctness of computed cycles]
If $f\colon A\to A$ is a bijection and $a\in \supp{f}$, then
$\cycle{f}{n}{a}$ is a prefix with head $a$, for all $n\in\mathbb{N}$. Moreover if $|\supp{f}| \leqslant n$, 
then $\cycle{f}{n}{a}$ is closed.

If $R$ is a list of prefixes and $as\subseteq\supp{f}$,
then $\cycles{f}{n}{as}{R}$ is a list of prefixes; if
$|\supp{f}| \leqslant n$, then $\cycles{f}{n}{as}{R}$
is a list of closed prefixes.
\end{lemma}

\begin{theorem}
If $f\colon A\to A$ is a bijection, then $R= \cycles{f}{|\supp{f}|}{\supp{f}}{[]}$ is a list
of closed prefixes. Therefore $\fromFP{\tofpUnS{R}}\,a = f\,a$, for all $a\in A$. 
\end{theorem}

Notice that a composition of transpositions might mention elements that are not in the
support of the induced permutation; for example, both $\swap 1 1$ and $\swap 1 2 \swap 2 1$ are equal
to the identity permutation. One can get a ``normalized'' representation by composing 
our functions. As a matter of fact, this was our motivation to formalize cycles.
\begin{corollary}[Normalization of transpositions] Let $p$ be a list of transpositions and
$\mathit{ats} = \supp{\tofpUn{p}}$. Moreover, let $R = \cycles{\tofpUn{p}}{|\mathit{ats}|}{\mathit{ats}}{[]}$.
Then $\fromFP{\tofpUnS{R}} = \fromFP{as}$; moreover every atom in $\tofpUnS{R}$ is in
its support.
\end{corollary}

\section{Our Formalization in Agda}
\label{sec:formalization}
\begin{figure}[ht]
  \centering
  \tikzset{every picture/.style={line width=0.75pt}} 

\begin{tikzpicture}[x=0.75pt,y=0.75pt,yscale=-1,xscale=1,scale=1]

\draw (260,20) node (A) [rectangle,draw,dashed,inner sep=10pt,rounded corners=4pt] {\agdainline{Algebra}};
\draw (160,90) node (GA) [rectangle,draw,inner sep=10pt,rounded corners=4pt] {\agdainline{GroupAction}};
\draw (360,90) node (P) [rectangle,draw,inner sep=10pt,rounded corners=4pt] {\agdainline{Permutation}};
\draw (260,160) node (N) [rectangle,draw,inner sep=10pt,rounded corners=4pt] {\agdainline{Nominal}};

\path [->,shorten <=2pt,shorten >=2pt,dashed] (GA) edge (A) ;
\path [->,shorten <=2pt,shorten >=2pt,dashed] (P) edge (A) ;
\path [->,shorten <=2pt,shorten >=2pt,dashed] (N) edge (P) ;
\path [->,shorten <=2pt,shorten >=2pt,dashed] (N) edge (GA) ;

\node [below=4pt of A.south,font=\small,xshift=-2ex] {Group};
\node [below=4pt of P.south east,font=\small,inner sep=0.75pt,align=left,xshift=-3ex] {Finite permutations\\Cycles\\Normalization\\$\Perm{\mathbb{A}}$ group};
\node [below=4pt of GA.south,font=\small,inner sep=0.75pt,align=left] {G-Sets\\Equivariant\\\hspace{2ex}functions\\Constructions};
\node [below=4pt of N.south,font=\small,inner sep=0.75pt,align=left] {Support\\Equivalence\\Constructions};

\end{tikzpicture}
  \caption{High level view of the modular organization in the project.}
  \label{fig:proj}
\end{figure}
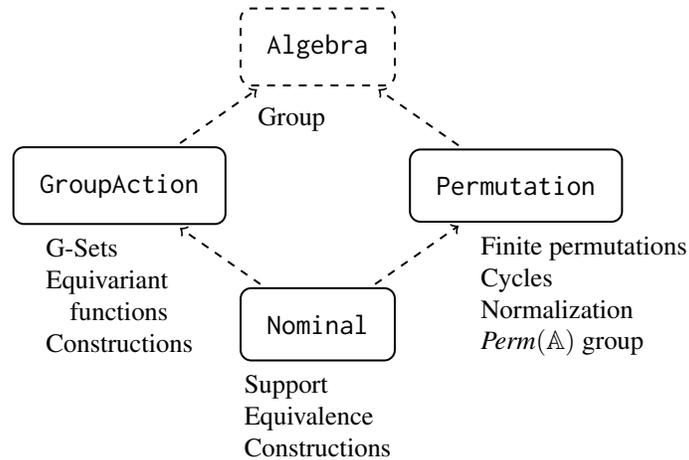

Our formalization is developed on top of the Agda's standard library
v1.7 \cite{agdalib}. Figure \ref{fig:proj} shows a high level view of
the project.  The standard library includes an algebraic hierarchy
going beyond groups; it lacks, however, a formalization of group
actions. The module \agdainline{GroupAction} includes G-Sets,
equivariant functions and constructions like products and co-products.
We also have a \agdainline{Permutation} module which includes the
concepts of finite permutations, cycles, normalization and the
permutation group.  And last, in the module \agdainline{Nominal} we
formalize the concepts of support, nominal set, equivalence between
different notions of support, normalization and again constructions
like products and co-products.

\noindent We first present the definition of Group in the standard library in order
to introduce some terminology and concepts:
\begin{minted}{agda}
    record Group c ℓ : Set (suc (c ⊔ ℓ)) where
      field
        Carrier : Set c
        _≈_     : Rel Carrier ℓ
        _∙_     : Op₂ Carrier
        ε       : Carrier
        _⁻¹     : Op₁ Carrier
        isGroup : IsGroup _≈_ _∙_ ε _⁻¹
\end{minted}
\noindent A Group is a \emph{bundle} where the components of its definition
(the carrier set, the unit, the inverse, the composition) are explicitly
mentioned plus a proof, given by \textit{isGroup}, that they satisfy the
axioms. Notice that one of the fields is a relation \textit{_≈_};
that relation should be an equivalence relation over the carrier:
essentially this amounts to say that the \agdainline{Carrier} has
a setoid structure. Setoids allows for greater flexibility
as they enable to work with a notion of equality that is not the 
propositional equality; \agdainline{Func X Y} is the set of functions
between setoids \agdainline{X} and \agdainline{Y} that preserve the
equality; sometimes these functios are called \emph{respectful}.

\paragraph{G-Sets}

Our first definition is the \emph{structure} that collects the equations
required for an action. In the following, we are under a module
parameterized by \mintinline{agda}{G : Group}.
\begin{minted}{agda}
    record IsAction (F : Func (G.setoid ×ₛ X) X) : Set _ where
      _●_ : Carrier G → Carrier X → Carrier X
      g ● x = Func.f F (g , x)
      field
        idₐ : ∀ x → ε ∙ₐ x ≈X x
        compₐ : ∀ g' g x → g' ● g ● x ≈X (g' ∙ g) ● x
\end{minted}
\noindent Notice that the record-type \mintinline{agda}{IsAction} is a 
predicate over respectuful functions from the setoid $G\times X$ to $X$.
The definition of G-Set is straightforward and follows the pattern of the
standard library
\begin{minted}{agda}
    record G-Set : Set _ where
      field
        set : Setoid ℓ₁ ℓ₂
        action : Func (G.setoid ×ₛ set) set
        isAction : IsAction action
\end{minted}
In order to introduce the notion of equivariant function\footnote{We note that
both Choudhury and Paranhos define equivariant functions only for the group of
finite permutations.} we first introduce the predicate
\agdainline{IsEquivariant} stating when \agdainline{H : Func X Y} is equivariant
for respectful functions \agdainline{FX} and \agdainline{FY}.
\begin{minted}{agda}
    IsEquivariant :
     {X : Setoid ℓ₁ ℓ₂} →
     {Y : Setoid ℓ₃ ℓ₄} →
     (FX : Func (G.setoid ×ₛ X) X) →
     (FY : Func (G.setoid ×ₛ Y) Y) →
     (H : Func X Y) → Set (ℓ₁ ⊔ ℓ₄ ⊔ cℓ)
    IsEquivariant {Y = Y} FX FY H = ∀ x g → F.f (g ●X x) ≈Y (g ●Y F.f x)
      where _●X_ = _●_ {F = FX} ; _●Y_ = _●_ {F = FY} ; _≈Y_ = _≈_ Y
            open module F = Func H
\end{minted}
\noindent Now we pack a respectful function between the setoids of G-Sets together with
a proof of it being equivariant.
\begin{minted}{agda}
    record Equivariant (X : G-Set) (Y : G-Set) : Set _ where
      field
        F : Func (set X) (set Y)
        isEquivariant : IsEquivariant (action X) (action Y) F
\end{minted}

In the following snippet we show how to
construct binary products of G-Sets (we use copatterns to define record objects).
\begin{minted}{agda}
    variable X Y : G-Set G
    private
      open module GX = G-Set X ; open module GY = G-Set Y
    G-Set-× : G-Set G
    set G-Set-× = GX.set ×ₛ GY.set
    f (action G-Set-×) (g , (x , y)) = g GX.● x  , g GY.● y
    cong (action G-Set-×) (g=g' , (x=x' , y=y')) =
      Func.cong GX.action (g=g' , x=x') ,  Func.cong GY.action (g=g' , y=y')
    idₐ (isAction (G-Set-×)) (x , y) = GX.idₐ x , GY.idₐ y
    compₐ (isAction (G-Set-×)) g g' (x , y) = GX.compₐ g g' x , GY.compₐ g g' y
\end{minted}
\noindent We now prove that the first projection is equivariant; notice that
\agdainline{G-Set-×} is the product of \agdainline{X} and \agdainline{Y} introduced
with the \agdainline{variable} keyword.
\begin{minted}{agda}
    π₁ : Equivariant G G-Set-× X
    f (F π₁) = proj₁
    cong (F π₁) = proj₁
    isEquivariant π₁ _ _ = refl (set X)
\end{minted}

\paragraph{Permutations}
Now we focus on the module \agdainline{Permutation}. We start by introducing the group $\SymA$
using the definitions of inverses from the standard library; notice that the equivalence 
relation is given by the point-wise (or extensional) equality of functions.
\begin{minted}{agda}
    -- In this context A-setoid is a Setoid (not necessarily decidable).
    A = Carrier A-setoid ; _≈A_ = _≈_ A-setoid
    Perm = Inverse A-setoid A-setoid
    _≈ₚ_ : Rel Perm _
    F ≈ₚ G = (a : A) → f F a ≈A f G a
  
    Sym : Group (ℓ ⊔ ℓ') (ℓ ⊔ ℓ')
    Carrier Sym = Perm
    _≈_ Sym = _≈ₚ_
    _∙_ Sym = _∘ₚ_        -- composition of Perm, from the stdlib
    ε Sym = idₚ A-setoid  -- identity Perm, from the stdlib
    _′ Sym = _⁻¹          -- inverse permutation, from the stdlib
    isGroup Sym = record { ... } -- ommited
\end{minted}

\noindent If we ask the setoid \agdainline{A-setoid} to be decidable, then we can define the swapping
permutation.
\begin{minted}{agda}
    module Perm (A-setoid : DecSetoid ℓ ℓ') where
      open DecSetoid A-setoid renaming (Carrier to A)
      transp : A → A → A → A
      transp a b c with does (c ≟ a)
      ... | true = b
      ... | false with does (c ≟ b)
      ... | true = a
      ... | false = c

      transp-perm : (a b : A) → Perm
      transp-perm a b = record { 
        f = transp a b ; f⁻¹ = transp a b
        ; cong₁ = transp-respects-≈ a b ; cong₂ = transp-respects-≈ a b
        ; inverse = transp-involutive a b , transp-involutive a b
    }
\end{minted}

Our next goal is to define the group $\PermA$ of finite permutations of atoms. As we explained
before, finite permutation can be given by a bijective map, as a composition of transpositions,
or as a composition of disjoint cycles. 

In other works the group of finite permutations is explicitly defined
as lists of pairs, where each pair represents a transposition and the
empty list is the identity permutation: appending a pair $(a,b)$ to a
list $p$ amounts to compose the transposition $\swap a b$ to the
permutation denoted by $p$. Concatenation of lists $p$ and $p'$ also
induces their composition. This choice has the advantage of being
explicit and avoids having alternative expressions for composing
permutations. On the other hand it still allows different
representatives for the same permutation; in fact, $[(a,a)]$,
$[(b,a),(a,b)]$, and $[]$ are all representations of \emph{the}
identity permutation. It is clear that the setoid of finite
permutations should equate those three versions of the identity,
therefore the equivalence relation used is that of inducing the same
permutation.

We started with the following syntactic representation of Finite Permutations, which is close
to that of lists but in terms of $S$-expressions; since we cannot ensure canonicity with lists,
why not to be more liberal also on associativity?
\begin{minted}{agda}
    data FinPerm : Set ℓ where
      Id : FinPerm
      Swap : (a b : A) → FinPerm
      Comp : (p q : FinPerm) → FinPerm
\end{minted}
The permutation associated with a \agdainline{FinPerm} is given by
\begin{minted}{agda}
    ⟦_⟧ : FinPerm → Perm
    ⟦ Id ⟧ = idₚ setoid
    ⟦ Swap a b ⟧ = transp-perm a b
    ⟦ Comp p q ⟧ =  ⟦ q ⟧ ∘ₚ ⟦ p ⟧
\end{minted}
Before introducing our concrete formalization of $\PermA$ let us exploit the fact that we 
have a decidable setoid of atoms to prove that the equivalence of finite permutation is 
also decidable. In order to do that, we define a relation \agdainline{_⊆ₛ_} on
\agdainline{FinPerm}; \agdainline{p ⊆ₛ q} holds when \agdainline{q} coincides with 
\agdainline{p} in the support of the latter. Since we can compute the support of
\agdainline{FinPerm}s and the equality of atoms is decidable, then we can decide 
\agdainline{_⊆ₛ_}.
\begin{minted}{agda}
    _⊆ₛ_ : Rel FinPerm (ℓ ⊔ ℓ')
    p ⊆ₛ q = All (λ a → f ⟦ p ⟧ a ≈ f ⟦ q ⟧ a) (support p)

    ?⊆ₛ : ∀ p q → Dec (p ⊆ₛ q)
    ?⊆ₛ p q = all? (λ a → f ⟦ p ⟧ a ≟ f ⟦ q ⟧ a) (support p)
\end{minted}
Moreover we can prove that the mutual containment is equivalent to denoting the same permutation;
thus we can decide the equality of finite permutations as given by \agdainline{FinPerm}:
\begin{minted}{agda}
    _≈ₛ_ : Rel FinPerm (ℓ ⊔ ℓ')
    p ≈ₛ q = p ⊆ₛ q × q ⊆ₛ p

    ≈ₛ-dec : ∀ p q → Dec (p ≈ₛ q)
    ≈ₛ-dec p q = (?⊆ₛ p q) ×-dec (?⊆ₛ q p)
    -- We omit the proofs of these lemmas.
    ≈ₛ⇒≈ₚ : ∀ p q → p ≈ₛ q → ⟦ p ⟧ ≈ₚ ⟦ q ⟧
    ≈ₚ⇒≈ₛ : ∀ p q → ⟦ p ⟧ ≈ₚ ⟦ q ⟧ → p ⊆ₛ q
    _≟ₚ_ : ∀ p q → Dec (⟦ p ⟧ ≈ₚ ⟦ q ⟧)
\end{minted}
Furthermore we can normalize a \agdainline{FinPerm} to have an equivalent permutation 
where every occuring atom is in its support. 

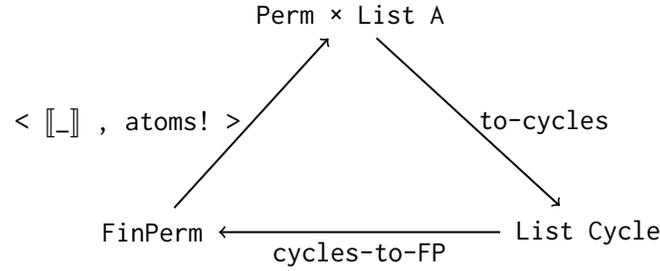
\begin{figure}
  \centering
  \tikzset{every picture/.style={line width=0.75pt}} 

\begin{tikzpicture}[x=0.75pt,y=0.75pt,yscale=-1,xscale=1]

\draw (260,15) node (P) {\agdainline{Perm × List A}};

\draw (380,125) node (C) {\agdainline{List Cycle}};

\draw (160,125) node (F) {\agdainline{FinPerm}};

\path [->,shorten <=2pt,shorten >=2pt] (P) edge node [right] {\agdainline{to-cycles}} (C); 
\path [->,shorten <=2pt,shorten >=2pt] (F) edge node [left] {\agdainline{< ⟦_⟧ , atoms! >}} (P); 
\path [->,shorten <=2pt,shorten >=2pt] (C) edge node [below] {\agdainline{cycles-to-FP}} (F); 

\end{tikzpicture}
  \caption{The mappings between different representations of permutations.}
  \label{fig:perms}
\end{figure}

Let us first revisit the Example \ref{ex:perm} 
now in Agda where we see how to encode a finite permutation as a composition of cycles.
\begin{minted}{agda}
    f : ℕ → ℕ
    f x with x ≤? 5
    ... | yes p = (x + 2) mod 6
    ... | no ¬p = x
\end{minted}
We represent cycles simply as lists of atoms; we certainly could also have used fresh lists to
represent cycles. A composition of cycles is a list of cycles.
\begin{minted}{agda}
    Cycle = List A
    
    cycle₀ cycle₁ : Cycle
    cycle₀ = 1 ∷ 3 ∷ 5 ∷ []
    cycle₁ = 0 ∷ 2 ∷ 4 ∷ []
    f-cycles : List Cycle
    f-cycles = cycle₀ ∷ cycle₁ ∷ []
\end{minted}
Or alternatively, it can also be expressed as a composition of four transpositions:
\begin{minted}{agda}
    f-swaps : FinPerm
    f-swaps = Comp (Comp (Swap 1 3) (Swap 3 5)) (Comp (Swap 0 2) (Swap 2 4))
\end{minted}
In Figure \ref{fig:perms} we show the three representations of finite permutations.
The normalization of \agdainline{FinPerm} is simply the composition of the mappings:
\begin{minted}{agda}
    norm : FinPerm → FinPerm
    norm = cycles-to-FP ∘ cycles-from-FP
\end{minted}
The functions \agdainline{cycles-to-FP} maps lists of disjoint cycles to
\agdainline{FinPerm} and \agdainline{cycles-from-FP} goes in the reverse direction,
producing a list of disjoint cycles from a \agdainline{FinPerm} (this is the
composition of the diagonal arrows in Fig.~\ref{fig:perms}).

The correctness of the normalization follows the proof presented in Sec.~\ref{sec:perm}.
Although we do not enforce neither freshness for cycles nor disjointness of cycles we
keep that as an invariant when we compute the cycles in \agdainline{to-cycles}.
\begin{minted}{agda}
    module Thm (p : FinPerm) where
      ats = atoms! p -- Fresh list of the atoms in the support of p.
      -- from-atom-~* is the proof of Lemma 3.
      rel = from-atoms-~* ⟦ p ⟧ ats []* (fp-supp p) (dom⊇atoms! p)
      -- the representation as composition of cycles
      ρs = to-cycles ⟦ p ⟧ (length ats) ats []
      -- This property follows from Lemma 3.
      ∈-dom⇒∈ρs : (_∈-dom ⟦ p ⟧) ⊆ (_∈ concat ρs)
    
      norm-corr : ⟦ p ⟧ ≈ₚ ⟦ norm p ⟧
      norm-corr x with x ∈? concat ρs
      ... | yes x∈at = ~*-out ⟦ p ⟧ rel x∈at      -- Item 3 of Lemma 1.
      ... | no x∉at = trans                       -- f ⟦ p ⟧ x = x = f ⟦ norm p ⟧ x
          (¬∈-dom⇒∉-dom {⟦ p ⟧} (contraposition ∈-dom⇒∈ρs x∉at))
          (~*-out-fresh ⟦ p ⟧ rel x∉at)           -- Item 4 of Lemma 1.
\end{minted}
We also have other correctness result to prove that the \agdainline{FinPerm}
obtained from a \agdainline{Perm} and its support is equivalent to it:
\begin{minted}{agda}
    module Thm’ (F : Perm) {ats : List A} (is-sup : ats is-supp-of F)
      (incl : (_∈ ats) ⊆ (_∈-dom F)) where
    
      ρs = to-cycles p (length ats) ats []
      norm-corr : F ≈ₚ ⟦ cycles-to-FP ρs ⟧
\end{minted}

Let us remark that \agdainline{FinPerm} is just a representation and the set of
finite permutation, \agdainline{PERM}, is the subset of \agdainline{Perm}
corresponding to the image of \agdainline{⟦_⟧}:
\begin{minted}{agda}
    PERM : Set _
    PERM = Σ[ p ∈ Perm ] (Σ[ q ∈ FinPerm ] (p ≈ₚ ⟦ q ⟧))
\end{minted}
A disadvantage of using this encoding is that we need to deal with triples; for instance, the
identity \agdainline{PERM} is represented by \agdainline{Id}.
\begin{minted}{agda}
    ID : PERM
    ID = idₚ setoid , Id , λ _ → refl
\end{minted}
The group $\PermA$ is explicity defined as:
\begin{minted}{agda}
    Perm-A : Group (ℓ ⊔ ℓ') (ℓ ⊔ ℓ')
    Carrier Perm-A = PERM
    _≈G_ Perm-A = _≈ₚ_ on proj₁
    _∙_ Perm-A = _∘P_
    ε Perm-A = ID
    _′ Perm-A = _⁻¹P
    isGroup Perm-A = record { ... }
\end{minted}

We alleviate the burden of working with triples by proving lemmas characterizing the action 
of \agdainline{PERM}s in terms of the finite permutation, for instance for \agdainline{Id}:
\begin{minted}{agda}
    -- In this context the group acting on G-Sets is Perm-A. 
    module Act-Lemmas {X-set : G-Set {ℓ₁ = ℓx} {ℓ₂ = ℓx'}} where
      _≈X_ = Setoid._≈_ set
      id-act : ∀ (π : PERM) (x : X) → proj₁ π ≈ₚ ⟦ Id ⟧ → (π ● x) ≈X x
      id-act π x eq = trans (congˡ {π} {ID} x eq) (idₐ x)
\end{minted}

\paragraph{Nominal Sets}
Remember that a subset $A\subseteq \mathbb{A}$ is a support for $x$ 
if every permutation fixing every element of $A$ fixes $x$, through the
action. A subset of a setoid \agdainline{A} can be defined either as a 
predicate or as pairs (just as in \agdainline{PERM} where the predicate 
is \agdainline{λ p → Σ[ q ∈ FinPerm ] (p ≈ₚ ⟦ q ⟧)}) or as another type, say
\agdainline{B}, together with an injection \agdainline{ι : Injection B A}.
\begin{minted}{agda}
    variable
      X : G-Set 
      P : SetoidPredicate A-setoid
    is-supp : Pred X _
    is-supp x = (π : PERM) → (predicate P ⊆ _∉-dom (proj₁ π)) → (π ● x) ≈X x
\end{minted}
The predicate \agdainline{λ a → f (proj₁ π) a ≈A a} is \agdainline{_∉-dom (proj₁ π)}; therefore, if \agdainline{P a} iff $a\in A$, then 
\agdainline{predicate P ⊆ _∉-dom (proj₁ π)} is a correct 
formalization of $\forall a\in A.\ π\,a=a$. 

Our official definition of support is the following:
\begin{minted}{agda}
    _supports_ : Pred X _
    _supports_ x = ∀ {a b} → a ∉ₛ P → b ∉ₛ P → SWAP a b ● x ≈X x
\end{minted}
Here \agdainline{SWAP} is a \agdainline{PERM}utation equal to 
\agdainline{⟦Swap a b⟧}. We formally proved that both definitions are equivalent, which is stated by the mutual implications:
\begin{minted}[baselinestretch=1]{agda}
    is-supp⊆supports : ∀ x → is-supp x → _supports_ x
    supports⊆is-supp : _supports_ ⊆ is-supp
\end{minted}
Let us note that the second implication uses explicitly the normalization of
finite permutations and its correctness.

In order to define nominal sets we need to choose how to say that a subset is
finite; as explained by Coquand and Spiwak \cite{Coquand2010} there are several
possibilities for this. We choose the easiest one: a predicate is finite if there
is a list that enumerates all the elements satisfying the predicate.
\begin{minted}{agda}
    finite : Pred (SetoidPredicate setoid) _
    finite P = Σ[ as ∈ List Carrier ] (predicate P ⊆ (_∈ as))
\end{minted}
A G-Set is nominal if all the elements of the underlying set are finitely 
supported.
\begin{minted}{agda}
    record Nominal (X : G-Set) : Set _ where
      field
        sup : ∀ x → Σ[ P ∈ SetoidPredicate setoid ] (finite P × P supports x)
\end{minted}

It is easy to prove that various constructions are nominals; for instance
any discrete G-Set is nominal because every element is supported by the empty
predicate \agdainline{⊥ₛ}:
\begin{minted}{agda}
    Δ-nominal : (S : Setoid _ _) → Nominal (Δ S)
    sup (Δ-nominal S) x = ⊥ₛ , ⊥-finite , (λ _ _ → S-refl {x = x})
      where open Setoid S renaming (refl to S-refl)
\end{minted}
We have defined \agdainline{G-Set-⇒ X Y} corresponding to the G-Set
of equivariant functions from \agdainline{X} to \agdainline{Y}; now we can
prove that \agdainline{G-Set-⇒ X Y} is nominal, again with \agdainline{⊥ₛ}
as the support for any \agdainline{F : Equivariant X Y}.
\begin{minted}{agda}
    →-nominal : Nominal (G-Set-⇒ X Y)
    sup (→-nominal) F = ⊥ₛ , ⊥-finite , λ _ _ → supported
      where supported : ∀ {a b} x → f ((SWAP a b) ∙→ F) x ≈Y f F x
\end{minted}

\section{Conclusion}
\label{sec:conclusion}
Nominal techniques have been adopted in various developments. We
distinguish developments borrowing some concepts from nominal
techniques to be applied in specific use cases (e.g. formalization of
languages with binders like the $\lambda$ or $\pi$ calculus with their
associated meta-theory) \cite{Bengtson2007, Copello2016, Copello2018,
Copello2018-2} from more general developments aiming to formalize at
least the core aspects of the theory of nominal sets. We are more
concerned with the later type.

The nominal datatype package for Isabelle/HOL \cite{Urban2006}
developed by Urban and Berghofer implements an infrastructure for
defining languages involving binders and for reasoning conveniently
about alpha-equivalence classes. This Isabelle/HOL package inspired
Aydemir et al.~\cite{Aydemir2007} to develop a proof of concept for
the Coq proof assistant, however it had no further development.  In
his Master thesis~\cite{Chou2015-mt}, Choudhury notes that none of the
previous developments following the theory of nominal sets were based
on constructive foundations.  He showed that a considerable portion
(most of the first four chapters of Pitts book \cite{Pitts2013-book})
of the theory of nominal sets can also be developed constructively by
giving a formalization in Agda. Pitts original work is based on
classical logic, and depends heavily on the existence of the smallest
finite support for an element of a nominal set. However, Swan
\cite{Swan2017} has shown that in general this existence cannot be
constructively guaranteed, as it would imply the law of the excluded
middle.

Choudhury works with the notion of \emph{some non-unique support}. In
order to formalize the category of Nominal Sets, Choudhury preferred
setoids instead of postulating functional extensionality. As far as we
know, Choudhury is still the most comprehensive mechanization in terms
of instances of constructions having a nominal structure.

Recently Paranhos and Ventura \cite{Paranhos2022} presented a
constructive formalization in Coq of the core notions of nominal sets:
support, freshness and name abstraction. They follow closely
Choudhury’s work in Agda \cite{Chou2015-mt}, acknowledging the
importance of working with setoids. They claim that by using Coq’s
type class and setoid rewriting mechanism, much shorter and simpler
proofs are achieved, circumventing the ``setoid hell'' described by
Choudhury. In his master thesis \cite{paranhos-thesis} Paranhos
further developed the library.

Both of those two formalizations in type theory take a very pragmatic
approach to finite permutations: a finite permutation is a list of
pairs of names. In our approach, we start with the more general
notion of bijective function from which the finite permutations are
obtained as a special case; moreover having different representations
allowed us to state and prove some theorems that cannot even be stated
in the other formalizations. So far, our main contributions are: the
representation of finite permutations and the normalization of
composition of transpositions; the equivalence between two definitions
of the relation ``$A$ supports the element $x$''; and proving that the
extension of every container type can be enriched with a group action
(notice that this cover lists, trees, etc.).

Our next steps are the definition of freshness. We are studying an
alternative notion of support that would admit having a freshness
relation between elements of two nominal sets (in contrast with other
mechanization that only consider ``the atom $a$ is fresh for $x$'')
and name abstraction. In parallel we hope to be able to prove that
extensions of finite containers on nominal sets are also nominal
sets. We also hope to streamline further some rough corners of our
development.

\subsection*{Acknowledgments} This formalization grew up from
discussions with the group of the research project ``Type-checking for
a Nominal Type Theory'': Maribel Fernández, Nora Szasz, Álvaro
Tasistro, and Sebastián Urciouli. We thank Cristian Vay for
discussions about group theory. This work was partially funded by
Agencia Nacional de Investigación e Innovación (ANII) of Uruguay.

\bibliographystyle{eptcs}
\bibliography{bibliography}
\end{document}